# Measuring the tuning curve of spontaneous parameter down-conversion using a comet-tail-like pattern

CHEN YANG,[1, 2] ZHI-YUAN ZHOU,[1, 2, 3] YAN LI,[1, 2] YIN-HAI LI,[1, 2] SU-JIAN NIU,[1, 2] ZHENG GE,[1, 2] GUANG-CAN GUO,[1, 2] AND BAO-SEN SHI [1, 2, 4]

[1] *CAS Key Laboratory of Quantum Information, University of Science and Technology of China, Hefei, Anhui 230026, China*
[2] *Synergetic Innovation Center of Quantum Information & Quantum Physics, University of Science and Technology of China, Hefei, Anhui 230026, China*
[3] *zyzhouphy@ustc.edu.cn;*
[4] *drshi@ustc.edu.cn*

**The comet-tail-like interference patterns are observed using photons from the spontaneous parametric down-conversion (SPDC) process. The patterns are caused by the angular-spectrum-dependent interference and the diffraction of a blazed grating. We present the theoretical explanation and simulation results for these patterns, which are in good agreement with the experimental results. The most significant feature of the patterns is the bright parabolic contour profile, from which, one can deduce the parameter of the parabolic tuning curve of the SPDC process. This method could be helpful in designing experiments based on SPDC.**

The spontaneous parametric down-conversion (SPDC) [1, 2], a second-order nonlinear process in a crystal without inversion symmetry, is widely used to produce both entangled photon pairs and heralded single photons. In this process, a higher energy pump photon splits into a pair of lower-energy photons, where one is designated a signal photon and the other an idler photon, and the twin photons possess non-classical temporal and spatial correlations [2, 3]. Because of the correlation properties, the quantum sources based on SPDC have been essential resources of many experiments that study fundamental problems and non-classical properties in quantum optics [4-7] and applications in quantum information, including quantum cryptography [8, 9], quantum teleportation [10, 11], quantum computation [12, 13], quantum sensing [14, 15] and quantum imaging [16, 17].

Except for the correlation properties, the photons from the SPDC process have a unique spectrum property: a structured frequency-angular spectrum caused by the phase-matching conditions [18]; the photons have an emission angle outside the crystal, and the frequency of the photons increases with increasing emission angle. For a long crystal, this dependence relation between frequency and emission angle is approximately a one-to-one mapping that is governed by a tuning curve and the tuning curve is helpful in designing experiments based on SPDC [19]. Our aim in this work is to present a new method to measure the tuning curve of SPDC.

In our previous work [20], we calculated the approximate parabolic expression for the tuning curve of a nondegenerate SPDC process and experimentally demonstrated an angular-spectrum-dependent (ASD) interference, which offers promise for applications in metrology. The fringe distribution of such interference is highly dependent on the expression of the tuning curve, therefore, Ref. [20] also provided a method to measure the tuning curve: by fitting the distribution of interference fringes. Besides, one can usually measure the tuning curve using a pinhole and a spectrometer. In this work, using a blazing grating, we observe comet-tail-like interference patterns and the "comet-tails" also have a parabolic contour profile, based on which one can directly deduce the expression of the parabolic tuning curve.

The experiment setup, as shown in Fig. 1, is based on the setup in our previous work [20], in which a 525 nm laser pumps a type-0 PPKTP crystal; 795 nm signal photons and 1540 nm idler photons are generated from the SPDC process. We set up a Michelson interferometer for the signal photons and discard the idler photons. Lens1 and Lens2 form a 4-f imaging system and an imaging plane of the crystal is located at the rear focal plane of Lens2. Lens2 and Lens3 also form a 4-f system and the field distributions on Mirror1 and Mirror2 are overlapped and imaged into the ICCD. The focal length of Lens1, Lens2, and Lens3 are 100 mm, 100 mm, and 200 mm respectively. Here, a grating is located at a position that is 100-mm away from the rear focal plane of Lens 2, this is the only difference from the setup in Ref. [20]. If the grating is a plane mirror, ring-like interference fringes can be recorded by the ICCD, which is the ASD interference presented in the Ref.[20], and the different frequency components are distributed radially and discretely in different rings, like the concentric rings shown in Fig. 2 (a). When reflected by a blazed grating, these frequency components redistribute horizontally through diffraction, and the larger the radius the larger the diffraction angle. The blue arrows in Fig. 2 (a) show the translation direction and their lengths represent the relative translation distances. The translation result, shown in Fig. 2 (b), is a group of non-concentric rings shaped like a comet tail. Fig. 2 (c) shows an ASD interference pattern without the grating inserted and Fig. 2 (d) shows a simulation result for the comet-tail-like pattern. Because the frequency components gather together on the left side in Fig. 2 (b), an extremely bright strip can be seen in actual interference patterns in Fig. 2 (d).

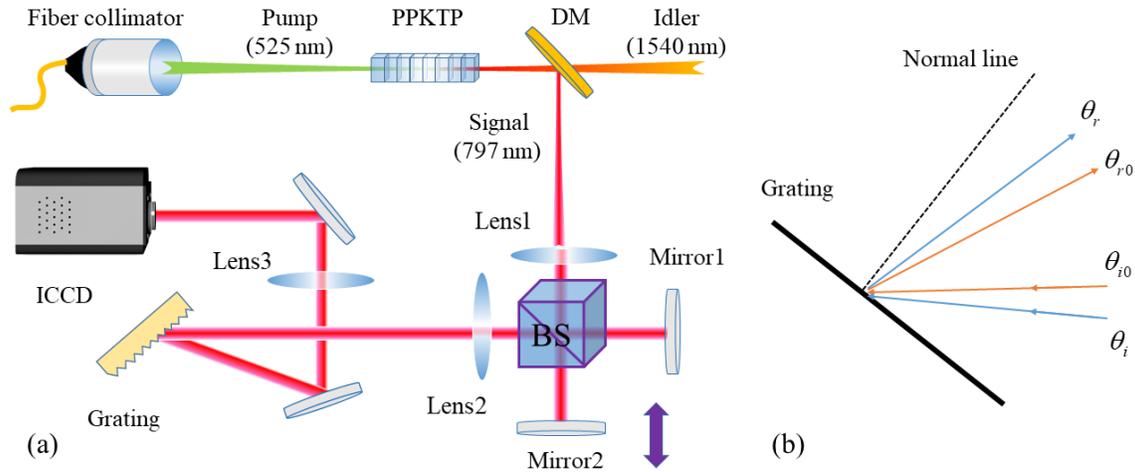

Figure 1. (a) A schematic of the experimental setup. PPKTP is the periodically poled potassium titanyl phosphate crystal which has a dimension of 1 mm × 2 mm × 10 mm and a grating period of 9.34 μm; DM: long-pass dichroism mirror; BS: beam splitter; ICCD: intensified charge-coupled device (Andor iStar DH334T); a half waveplate and a quarter waveplate for 525 nm are omitted which are used to transform the pump beam from the collimator into a vertically polarised beam; a 750 nm long-pass filter located before Lens1 is omitted which is used to filter the pump beam. (b) The diagram of the incident and reflecting angles around the grating. $\theta$ represents the angle between the normal line and the ray of light.

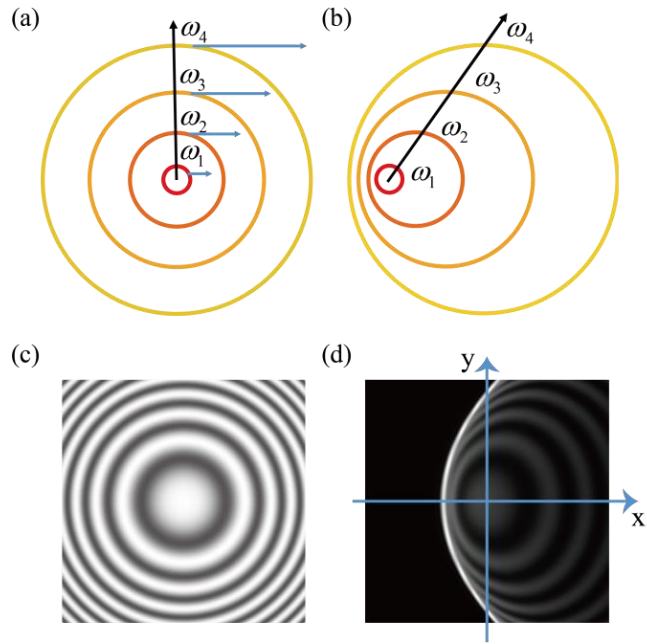

Figure 2. (a)/(b) The frequency distribution of the signal photons before/after the grating; (c)/(d) Simulation results with an arm difference of 100 mm before/after the grating being inserted.

In the following, we first explain how the simulation is performed. To obtain the simulation result for the comet-tail-like interference pattern in Fig. 2 (d), every pixel value of the original ring-like pattern in Fig. 2 (c) (simulation method and more data are shown in Ref. [20]) is translated horizontally using the grating equation

$$d_g(\sin\theta_{in} + \sin\theta_r) = \lambda_s, \tag{1}$$

where $\theta_{in}$ and $\theta_r$ represent the incident and reflecting angle of the light with a signal photon wavelength of $\lambda_s$; $d_g$ represents the grating constant that is $1/1200$ mm in our experiment. To describe how to translate the interference patterns, we assume an x-y coordinate system (shown in Fig. 2 (d)) on the detection plane, where the component of center wavelength $\lambda_{s0}$ is detected at the origin. The constant $\theta_{in0}$ and $\theta_{r0}$ represent the incident and reflecting angle for the light with the center wavelength $\lambda_{s0}$, and the relationship between these angles are shown in Fig. 1 (b). The incident angle in our experiment is of order $40°$ and the reflecting angle can be solved using the equation $d_g(\sin\theta_{in0} + \sin\theta_{r0}) = \lambda_{s0}$. The coordinates are then given by

$$x = (\theta_r - \theta_{r0})f \tag{2}$$

$$x' = (\theta_{in} - \theta_{in0})f = \pm\sqrt{f^2\theta_{s\_out}^2 - y'^2} \tag{3}$$

$$y = y' \tag{4}$$

where $(x', y')$ and $(x, y)$ represent the coordinate before and after the translation, $f$ represents the focal length of Lens3, $\theta_{s\_out}$ represents the outside emission angle of signal photons, and $f\theta_{s\_out} \approx \sqrt{x'^2 + y'^2}$ gives the radius of the ring patterns.

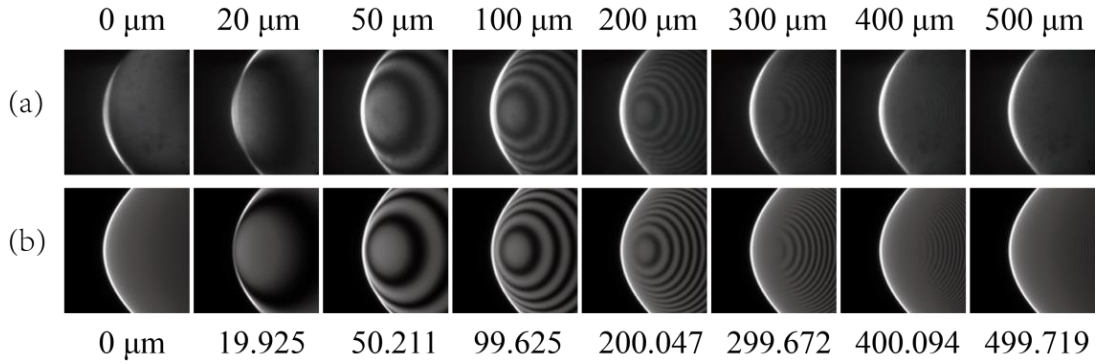

Figure 3. Comet-tail-like interference patterns. (a) Experimental results. (b) Simulation results.

In the simulation, the light wavelength is discretized; first, we simulate the interference results for each discretized wavelength component; then, we translate them using Eq. (1) ~ (4); finally, we sum the translation results of all the wavelength components and the interference patterns like that in Fig. 2 (d) are obtained.

In the experiment, we fixed Mirror2 on a displacement platform and Mirror1 on a piezoelectric transducer. We varied the arm difference by moving Mirror2, then finely adjusted the PZT to ensure the centers of the interference patterns are bright spots. Fig. 3 (a) and (b) show the experimental results and more simulation results respectively. The numbers at the top indicate approximate arm differences, specifically, from the reading of the displacement platform. The numbers at the bottom of Fig. 3 (b) show the actual arm differences set in the simulation. In comparison, the experimental results agree well with our theoretical calculation. The most striking feature of the comet-tail-like interference patterns is their bright parabolic contour profile. We then give the derivations of the approximate parabolic expression to explain its cause and establish its quantitative relation with the expression for tuning curve, which is given by the equation [20]

$$\theta_{s\_out}^2 = b_1\Delta\omega = -b_2\Delta\lambda \tag{5}$$

$$b_1 = \frac{2n_{i0}n_{s0}\omega_{i0}(\beta_s\omega_{s0} + n_{s0} - \beta_i\omega_{i0} - n_{i0})}{\omega_{s0}(\omega_{s0}n_{s0} + \omega_{i0}n_{i0})} \tag{5a}$$

$$b_2 = (2\pi c / \lambda_{s0}^2) b_1 \tag{5b}$$

where $\Delta\omega = \omega_s - \omega_{s0}$, $\Delta\lambda = \lambda_s - \lambda_{s0}$ and $\beta_s = \left(\dfrac{dn}{d\omega_s}\right)_{\omega_{s0}}$, $\beta_i = \left(\dfrac{dn}{d\omega_i}\right)_{\omega_{i0}}$ are the coefficients of first-order dispersion at the center frequency of the signal and idler photons. The subscript 0 indicates a corresponding value at the center frequency of the wavelength. The following derivations are based on Eq. (1) and (5) and the aim is to describe the parabolic stripe using the parameter $b$. Performing the differential operation for Eq. (1), we have

$$\cos\theta_{in} d\theta_{in} + \cos\theta_r d\theta_r = 0. \tag{6}$$

Because the divergence angle of photons is relative small, the differentials here can be approximately written as

$$d\theta_{in} \approx \theta_{in} - \theta_{in0} = \pm\sqrt{\theta_{s\_out}^2 - (y/f)^2}, \tag{7}$$

$$d\theta_r \approx \theta_r - \theta_{r0} \approx x/f, \tag{8}$$

Next, from Eq. (5) ~ (8), one can obtain the horizontal coordinate

$$x(\lambda_s, y) = -\dfrac{\cos\theta_{i0}}{\cos\theta_{r0}}\sqrt{b_2(\lambda_{s0} - \lambda_s)f^2 - y^2} \pm \dfrac{(\lambda_{s0} - \lambda_s)f}{d_g \cos\theta_{r0}}. \tag{9}$$

To explain how the bright parabolic stripe generates, we plot the $x - \lambda_s$ curve with $y = 0$ and show it in Fig. 4 (a). On the left of the curve, most wavelength components gather together around $x = -2.4$ mm, therefore, an extremely bright point is formed after the auto integration over wavelength by the ICCD and is much brighter than elsewhere. The extremely bright point occurs where the partial derivative with respect to the wavelength is zero, then we have

$$\dfrac{\partial x}{\partial \lambda_s} = \dfrac{b_2 f^2 \cos\theta_{in0}}{2\cos\theta_{r0}\sqrt{b_2(\lambda_{s0} - \lambda_s)f^2 - y^2}} - \dfrac{f}{d_g \cos\theta_{r0}} = 0. \tag{10}$$

Substituting Eq. (10) into Eq. (9) and eliminating the variable $\lambda_s$, one can obtain the parabolic expression

$$x = ay^2 - c. \tag{11}$$

$$a = 1/(d_g b_2 f \cos\theta_{r0}) \tag{11a}$$

$$c = -d_g b_2 f \cos^2\theta_{in0} / (4\cos\theta_{r0}) \tag{11b}$$

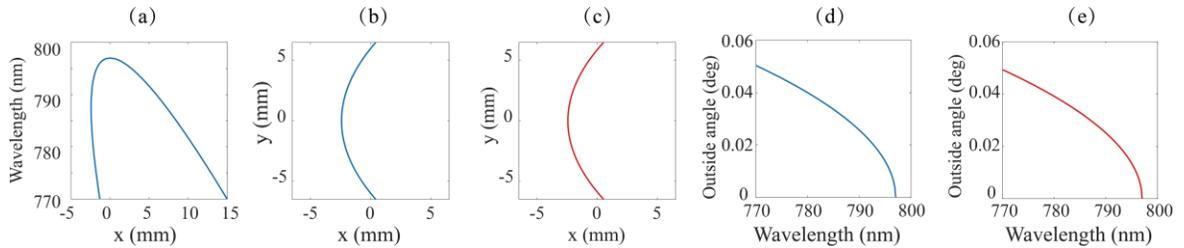

Figure 4. (a) The wavelength-position curve based on Eq. (9). (b) The parabolic curve plotted based on Eq. (11). (c) The parabolic curve plotted by fitting the bright parabolic stripe in Fig.2 (d). (d) The predictive tuning curve plotted based on Eq. (5) with $b_2 = 0.094\ \mu m^{-2}$; (e) The measured tuning curve that is plotted based on Eq. (5) with $b_2 = 0.090\ \mu m^{-2}$.

We plot the curve and show it in Fig. 4 (b). Its shape is in good agreement with that of the parabolic stripe in Fig. 2 (c) and Fig. 3 (b). The maximum pixel values in Fig. 2 (d) is fitted using a parabolic curve $x = a'y^2 - c'$ that is shown in Fig. 4 (c). Both $a'$ and $c'$ can be used to deduce the parameter $b_2$. Based on Eq. (11a) and (11b), we have $b_2 = 0.090 \pm 0.005\ \mu m^{-2}$ and $b_2 = 0.095 \pm 0.008\ \mu m^{-2}$ respectively; the predictive value calculated using Eq. (5a) and (5b) is $0.094\ \mu m^{-2}$. The error is mainly from the width of the parabolic stripe that can be seen in Fig. 3 (b). The measurement based on parameter $c$ has a larger uncertainty because the measurement error of incident angle $\theta_{in0}$ has a larger contribution on parameter $c$. With the parameter $b_2$, one can plot the tuning curves using Eq. (5). The predictive and measured tuning curve are shown in Fig. 4 (d) and (e) respectively, from which one can conclude that the measurement result agrees well with the predictive result.

Another interesting phenomenon in Fig. 3 is that the coherence length is shorter on the left side and longer on the right side of the "comet-tail". The reason is that the coherence length is in inverse proportion to the linewidth: from Fig. 4 (a), one can see that $\frac{d\lambda_s}{dx}$ on the left side is much larger than the right side, which means that the wavelength components gather together on the left whereas scatter on the right, therefore the linewidth at a spatial point on the left side is wider than the right side.

In the scenario of measuring the tuning curve, the Michelson interferometer is in fact not necessary. The expression for the parabolic contour profile is independent of the arm difference and as shown in Fig. 3, the parabolic contour profile can still occur when the path difference exceeds the coherence length. Here, the interferometer is just used to discretize the frequency components and show the principle clearly.

In summary, using a blazed grating, we observed comet-tail-like interference patterns based on the ASD interference. We have explained their cause and the simulation results are in good agreement with the experimental results. The most significant feature is the bright parabolic contour profile and from its expression, one can deduce the parameter of the tuning curve.

**Funding.** National Natural Science Foundation of China (NSFC) (11934013, 92065101); Anhui Initiative In Quantum Information Technologies (AHY020200).
**Disclosures.** The authors declare no conflicts of interest.
**Data availability.** Data underlying the results presented in this paper are not publicly available at this time but may be obtained from the authors upon reasonable request.